# Fabrication of Nanostructured PLGA Scaffolds Using Anodic Aluminum Oxide Templates

Cheng-Chih Hsueh[1], Gou-Jen Wang[*,1,2], Shan-hui Hsu[2,3], Huey-Shan Hung[3],
[1]Department of Mechanical Engineering, [2]Institute of Biomedical Engineering,
[3]Department of Chemical Engineering, National Chung-Hsing University,
Taichung 402, Taiwan

*Abstract-* PLGA (poly(lactic-*co*-glycolic acid)) is one of the most used biodegradable and biocompatible materials. Nanostructured PLGA even has great application potentials in tissue engineering. In this research, a fabrication technique for nanostructured PLGA membrane was investigated and developed. In this novel fabrication approach, an anodic aluminum oxide (AAO) film was use as the template; the PLGA solution was then cast on it; the vacuum air-extraction process was applied to transfer the nano porous pattern from the AAO membrane to the PLGA membrane and form nanostures on it. The cell culture experiments of the bovine endothelial cells demonstrated that the nanostructured PLGA membrane can double the cell growing rate. Compared to the conventional chemical-etching process, the physical fabrication method proposed in this research not only is simpler but also does not alter the characteristics of the PLGA. The nanostructure of the PLGA membrane can be well controlled by the AAO temperate.

## I. INTRODUCTION

Nanotechnology refers broadly to the use of materials and systems whose unifying subject is the control of matter at the atomic, molecular and supramolecular levels, and the fabrication of devices ranging from 1 to 100 nm. Nanobiotechnology is the branch of nanotechnology that has biological and biochemical applications or uses. Nanobiomaterial is one of the crucial fields in the study of nanobiotechnology. The major applications of nanobiomaterial are the fabrication of scaffolds for tissue engineering [1][2] and carriers for drug delivery [3-6]. The former refers to the fabrication of nanostructured patterns on a biomaterial through chemical processes. The efficacies of the subsequent cell culture and tissue growth using the nano patterned biomaterial can thus be enhanced; the latter involves the fabrication of nano-scale biodegradable materials as carriers to deliver a particular drug to a specific tissue.

For the applications of biomaterials in tissue engineering, the scaffolds eventually have to be implanted in a live body. Accordingly biodegradability and biocompatibility are the main considerations. Among those commercially available biomaterials, PLGA has been successfully utilized as a biodegradable material because it undergoes hydrolysis in the body to produce the original monomers, lactic acid and glycolic acid, and hence gives rise to very minimal systemic toxicity. Nano patterned PLGA even has great application potentials in tissue engineering. Savaiano and Webster [7] investigated long-term functions and intracellular responses of chondrocytes on novel nanophase PLGA/titania composites, indicating that nanostructured PLGA/nanophase titania composites possess properties to alter chondrocyte responses. Park et al. [8] used fabricated 2D and 3D NaOH-treated PLGA scaffolds and demonstrated that NaOH-treated PLGA scaffolds enhanced chondrocyte functions compared to non-treated scaffolds. Webster et al. [9] compared the adhesion of pseudomonas fluorescens on nanophase with that on conventional grain size alumina substrates. Results showed that nanostructured surface feature is the key material property to enhance bacterial adhesion. Miller et al. [10] formulated highly-controllable, nanostructured features on PLGA surfaces using a mica mold that was patterned by different sizes of polystyrene nanospheres. Results demonstrate that PLGA with nanometer spherical features promoted fibronectin spreading and subsequent vascular cell adhesion. Min et al. [11] employed the simultaneous electrospinning process to fabricate a composite matrix with chitin nanoparticles being embedded within a PLGA nanofiber matrix. Results indicated that the PLGA and PLGA/chitin matrices are good matrices for normal human fibroblasts.

The published studies regarding the fabrication of nanostructured features on PLGA surfaces can be categorized into the NaOH based chemical etching process and the physical molding process using a nanoparticle patterned replica. The chemical etching process encounters the problems of disorderly nanostructure and NaOH remaining. The physical molding process however requires more complicated procedures. The objective of the present study was to present a simply physical process for the fabrication of orderly nanostructured PLGA membranes. In this novel fabrication approach, an anodic aluminum oxide (AAO) film was use as the template; the PLGA solution was then cast on it; the vacuum air-extraction process was applied to transfer the nano porous pattern from the AAO membrane to the PLGA membrane and form nanostructures on it. The pattern and size of the nanostructures can be well controlled through the AAO template.

## II. MATERIALS AND METHODS

### 2.1 Materials

The physical processes involve in this study are the PLGA solution casting and the vacuum air-extraction. An AAO membrane was employed as the template for casting. A prepared PLGA solution was then cast on it. After the solvent was completely evaporated, the sample was put in a vacuum oven. The vacuum air-extraction was subsequently conducted to suck the semi-congealed PLGA into the nanopores of the AAO membrane. Stripping off the AAO template, orderly nanorods of PLGA could be obtained. In this manner, the preparation of materials includes the fabrication of the AAO templates with different pore diameters and the producing of the suitable PLGA solution.

*A. AAO film preparation*

The AAO films were prepared by using the well known anodizing process. Aluminum foils (99.9995%, 175 μm thick) were cleansed using ethanol and acetone in sequence, followed by annealing at 400°C for 3 hours in a vacuum. Electropolishing was then applied by using a sulfuric acid (96 wt %) and phosphoric acid (85 wt %) mixing solution as the electrolyte under a constant voltage of 20 V at 40 °C for 14 min to further polish the surfaces of the foil. An AAO film with nanopore diameter around 50 nm and thickness of 20 μm was obtained by anodizing aluminum foils in a 0.3 M phosphoric acid solution under 140 V of applied voltage at 0°C for 10 hours. An oxalic acid anodization process was also conducted under the conditions of 60 V of applied voltage at 0°C for 4 hours to obtain an AAO film with nanopore diameter around 20 nm. The remaining aluminum beneath the barrier layer was dissolved in an aqueous $CuCl_2 \bullet HCl$ solution that was prepared by dissolving 13.45 g powder of $CuCl_2$ into a 100 ml of 35 wt% hydrochloric acid solution. Following, immerse the bottom surface into the 30 wt% phosphoric acid for 120 minutes to etch out the barrier layer.

Figure 1 shows the SEM morphology of the phosphoric acid and oxalic acid processed AAO films. The nanopore diameters are around 50nm and 20nm, respectively. The AAO films were available as the templates for the vacuum air-extraction.

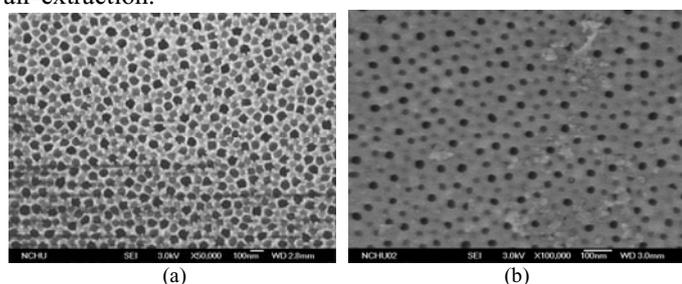

Fig1. SEM images of the AAO films; (a) Phosphoric acid processed (b) Oxalic acid processed

*B. PLGA solution preparation*

The PLGA solution was prepared by dissolving 85/15 POLY, IV：1.6-1.99 (dl/g), Mw:350000-500000 (Da) (Bio Invigor Corp., Taiwan) in acetone in a 1:4 w/w ratio, by stirring the mixture with a magneto agitator at 60°C for 60 min. The PLGA solution was then shaken by an ultrasonic shaker for 15 min to remove bubbles created during mixing. Others solvents such as chlorinated solvents, tetrahydrofurn, acetone or ethyl acetate, and dioxane can also be implemented to dissolve PLGA.

### 2.2 Orderly nanostructure imprinting

Figure 2 schematically illustrates the nanostructure imprinting scheme using an AAO template. The PLGA solution was cast on the AAO template, lasting for 1-2 hours until the solvent was completely evaporated and the surface had solidified. Following, the sample was placed in an oven undergoing the vacuum air-extraction with conditions of vacuum pressure being 76cm-hg at 40°C for 24 hours. The semi-congealed PLGA was sucked into the nanpres of the AAO template to form a nanorods array of PLGA.

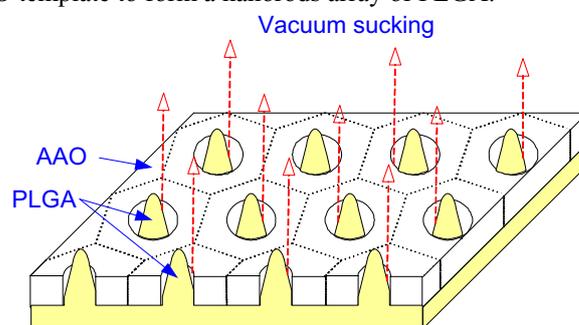

Fig. 2. Schematic illustration of the nanostructure imprinting scheme

Phosphoric acid can be applied to etch off the AAO template. However, we found that the PLGA membrane could be easily separated from the AAO template in cold water. Specifically, the sample was immersed in cold water for 2-3 minutes and then the PLGA membrane was peeled off. Figure 3 displays the AFM image of a nanopatterned PLGA membrane using an oxalic acid processed AAO template. The AFM image of a nanopatterned PLGA membrane produced by a phosphoric acid processed AAO template is shown in Figure 4. From both images, we can easily observe the nanostructures on the PLGA membrane.

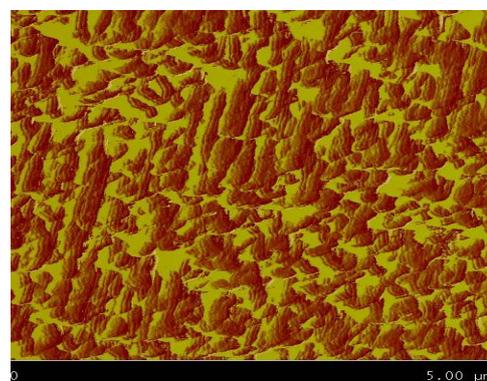

Fig. 3. AFM image of a nanopatterned PLGA membrane by an oxalic acid processed AAO template

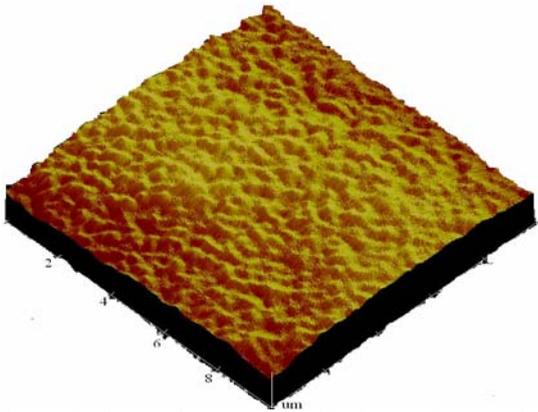

Fig. 4. AFM image of a nanopatterned PLGA membrane produced by a phosphoric acid processed AAO template

### III    CELL SEEDING

#### 3.1 Cells

Bovine endothelial cells (BEC) were used in the experiments to examine the adhesion capability of the presented nanostructured PLGA membrane. Cells were cultured in standard cell culture conditions (i.e., a 37°C, humidified, 5% $CO_2$/95% air environment).

#### 3.2 Cell density

Bovine endothelial cells were seeded at a density of $2\times10^4$ cells/well onto the presented nanopatterned PLGA scaffolds (phosphoric acid processed AAO template produced) and were cultured for 24 and 48 hours. For comparisons, BECs were also seeded on a flat PLGA scaffold. Experimental results shown in Figure 5 indicate that there were almost two-fold of cells on the presented nanopatterned PLGA scaffold after 24 and 48 hours of culture. Results of the present study demonstrated the exhilarating adhesion capability of the presented nanostructured PLGA membrane.

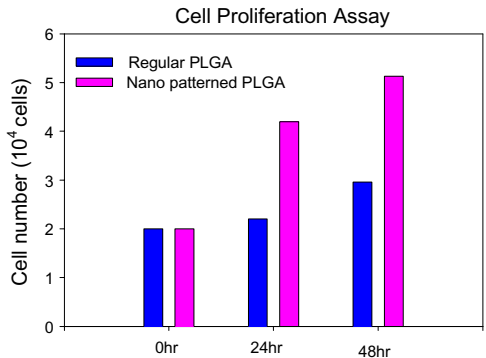

Fig. 5. The cell adhesion test

#### 3.3 Cell cytoskeleton labeling

The major challenge in observing the seeding cells on a PLGA scaffold is that the PLGA scaffold progressively turned opaque due to absorption of the cultivation medium into the PLGA and hydrolysis of the PLGA. Commonly, a PLGA scaffold becomes completely milky and opaque within 24 hours, making it difficult to monitor the progress of cell seeding by an optical microscopy (OM). To overcome the inability to effectively monitor the seeding cells, a suitable labeling process has to be conducted. The labeling procedures adopted in this study are briefly described as follows.

(i) Cell fixing: Gently rinsed the PLGA scaffold with PBS to remove any nonadherent cells. The adherent cells were then fixed with a 4% Paraformaldehyde (Sigma, P-6148, US) for 15 min. The sample was rinsed three times with PBS to completely wash off the culture medium.

(ii) Cell punching: The sample was then putting into reactions with a 0.5％ Triton X-100 (Sigma,T-9284) for 10 min at room temperature. Aspirate the mixing solution from the well followed by rinsing the well three times with PBS. Holes had been punched on cells, enabling stain to penetrate into the cytoplasm.

(iii) Mixed a 1% BSA (bovine serum albumin；United States Biochemical, USB, 10857, US) with the PBS. Injected the mixing solution into the well and processed the blocking reaction for 30 min.

(iv) Added stain phalloidin (1: 1000, Sigma) to the well and lasted for 30 min.

(v) The last step was pouring a preservation solution into the well and wrapped the well with an aluminum film. The sample was ready for further examination.

The fluorescence microscopy (Eclipse 80i, Nikon) images of the cytoskeleton labeling experiments for the flat PLGA scaffold, oxalic acid processed AAO template imprinted PLGA scaffold, and phosphoric acid processed AAO template imprinted PLGA scaffold are shown in Figure 6-8, respectively. BECs could adhere and grow healthily on those there PLGA scaffolds, especially on the scaffold imprinted by a phosphoric acid processed AAO template.

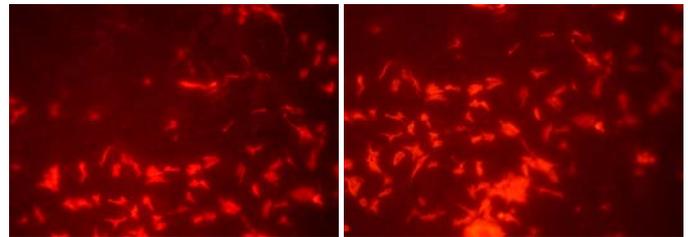

Fig. 6. Fluorescence microscopy image of the cytoskeleton labeling experiments for the flat PLGA scaffold

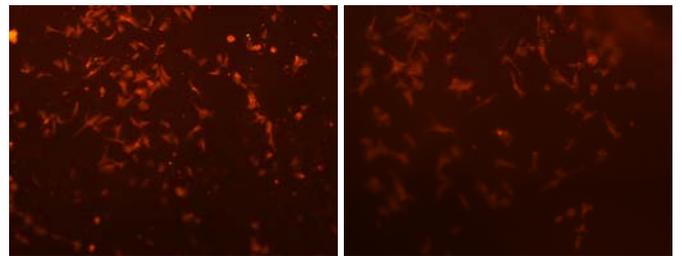

Fig. 7. Fluorescence microscopy image of the cytoskeleton labeling experiments for the oxalic acid processed AAO template imprinted PLGA scaffold

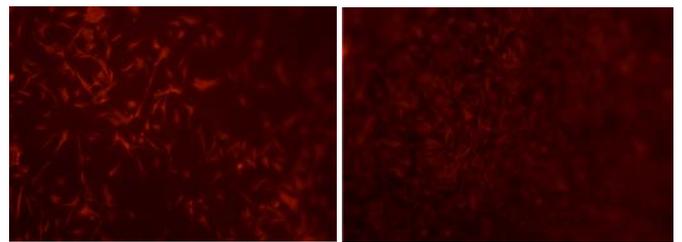

Fig. 8. Fluorescence microscopy image of the cytoskeleton labeling experiments

for the phosphoric acid processed AAO template imprinted PLGA scaffold

## IV. Conclusion

A simply physical process for the fabrication of orderly nanostructured PLGA membranes was presented in this study. An anodic aluminum oxide (AAO) film was use as the template together with the vacuum air-extraction process to imprint orderly nanostructures on a PLGA substrate. The cell culture experiments of the bovine endothelial cells demonstrated that the nanostructured PLGA membrane can largely increase the cell growing rate. Compared to the conventional chemical-etching process, the physical fabrication method proposed in this research not only is simpler but also does not modify the characteristics of the PLGA.